\documentclass[doublecol,amsmath,amssymb]{epl2} 
\usepackage{mathbbol}
\usepackage{graphicx}
\usepackage{color} 
\usepackage{cancel}
\usepackage{ulem}
\usepackage{soul}
\usepackage{bbold}

\newcommand{\ket}[1]{|#1\rangle}
\newcommand{\bra}[1]{\langle #1|}
\definecolor{orange}{rgb}{1,0.5,0}
\definecolor{gray}{rgb}{0.5,0.5,0.5}
\newcommand{\fer}[1]{#1}

\title{Orthogonal measurements are {\it almost} sufficient for quantum discord of two qubits}
\shorttitle{Orthogonal measurements are almost sufficient for quantum discord} 

\author{F. Galve\inst{1} \and G. L. Giorgi\inst{1} \and R. Zambrini\inst{1}}
\shortauthor{F. Galve \etal}

\institute{                    
  \inst{1} IFISC (UIB-CSIC), Instituto de F\'isica Interdisciplinar y Sistemas Complejos, UIB Campus,
E-07122 Palma de Mallorca, Spain
}

\abstract{
The common use in literature of orthogonal measurements in obtaining quantum discord for two-qubit states is discussed and compared with more general measurements. We prove the optimality of orthogonal measurements for rank 2 states. While for rank 3 and 4 mixed states they are not optimal, we present strong numerical evidence showing that they give the correct quantum discord up to minimal corrections. Based on the connection, through purification with an ancilla, between discord and entanglement of formation (EoF), we give a tight upper bound for the EoF of a $2\otimes N$ mixed state of rank 2, given by an optimal decomposition of 2 elements. We also provide an alternative way to compute the quantum discord for two qubits based on the Bloch vectors of the state.
}

\begin{document}

\maketitle

\section{Introduction}

The discovery and analysis of the quantum speedup within algorithms where entanglement is totally absent \cite{info-no-ent} has raised considerable interest to alternative measures of quantum correlations beyond entanglement. In contrast to the paradigm of a tensor product structure as the root of classicality of correlations, a new paradigm based on the ignorance produced by measurement has crystallized into several measures of quantum correlations, of which perhaps the most widely used is the quantum discord \cite{zurek,vedral}. In a sense, it captures the fact that unless measurements on party $B$ leave unaffected party $A$ in a bipartite states $\varrho_{AB}$, we cannot really speak of such state as being purely classically correlated.

An intense recent research activity is based on quantum discord as a  quantifier for quantum correlations for
two-qubit states. Yet it is typically used in a simplified form, where  only orthogonal measurements are
considered (see e.g. \cite{zurek,info-no-ent,maziero,luo2008,mazhar2010,mazzola2010,guo}). There is now a raising
concern about the possibility that more general measurements might modify the value of quantum discord, thus
weakening the conclusions of some recent works. 
The work by Hamieh et al. \cite{zaraket} took a first step showing  the sufficiency of projective measurements
(i.e. rank 1 POVM's) for states of two qubits. This of course does not demonstrate that two orthogonal  projectors
are enough. In spite of the great reduction of the complexity of the problem \cite{zaraket}, the optimal
projective measurement of a qubit can  have between 2 and 4 elements \cite{dariano} (the case of 2 elements
corresponds to orthogonal measurements) and the question of how many elements the optimal POVM has is still open.
In fact in the work by Hamieh et al. \cite{zaraket} only a very particular state is studied. Hence, the matter
remains unsettled and using orthogonal measurements seems to be an unnecessary restriction.

In this Letter, we show that orthogonal measurements are sufficient to obtain the discord of rank 2 states of two qubits, while for rank 3 and 4 they give a pretty tight upper bound. Moreover, \fer{given the relationship} between quantum discord and entanglement of formation \fer{ for pure tripartite states} \cite{winter}, we give a formula for discord based purely on the eigenvectors and eigenvalues of the original state, valid exactly for rank 2. We also show that the entanglement of formation of a $2\otimes N$ rank 2 state has a tight upper bound given by optimal decompositions of 2 elements. Finally, we give an alternative formula for quantum discord of two qubits states of any rank based on its Bloch vectors.

\section{Quantum discord}
Two classically equivalent formulas for the mutual information in a bipartite state, related by Bayes rule are $\mathcal{I}(A:B)=H(A)+H(B)-H(A,B)$ and $\mathcal{J}(A:B)=H(A)-H(A|B)$, where $H(.)$ is the Shannon entropy and $H(A|B)$ is the conditional Shannon entropy of $A$ given $B$. Their quantum counterparts, however, differ substantially \cite{zurek}, the former being known as the quantum mutual information:
\begin{equation}
\mathcal{I}(\varrho)=S(\varrho_A)+S(\varrho_B)-S(\varrho),
\end{equation}
where $S(.)$ is the von Neumann entropy and $\varrho_{A(B)}$ are the reduced states after tracing out party $B(A)$. It is precisely in $\mathcal{J}(A:B)$ where measurements come into scene, since the conditional entropy of {\it A given B} is the one given by measurement outcomes on party B. Though a general measurement has to be associated to a generic POVM, for the sake of simplicity the community has used the restricted set of perfect (von Neumann, or orthogonal) measurements, i.e. :
\begin{equation}
{\cal{J}}(\varrho)_{\{\Pi_j^B\}}=S(\varrho_A)-S(A|\{\Pi_j^B\})
\end{equation}
 with the conditional entropy defined as $S(A|\{\Pi_j^B\})=\sum_ip_iS(\varrho_{A|\Pi_i^B})$, $p_i={\mbox
Tr}_{AB}(\Pi_i^B\varrho)$ and where $\varrho_{A|\Pi_i^B}= \Pi_i^B\varrho\Pi_i^B/{p_i} $ is the density
matrix after a complete projective measurement $(\{\Pi_j^B\})$ has been performed on B. Notice that $\Pi_j^B$ are orthogonal projectors. Quantum discord is thus defined as the difference 
\begin{equation}
\label{eqdisc}
\delta_{A:B}(\varrho)=\min_{\{\Pi_i^B\}}\left[S(\varrho_B)-S(\varrho)+S(A|\{\Pi_i^B\})\right],
\end{equation}
minimized over all possible orthogonal measurements.
However, more general measurements should be used to exhaust the minimization problem, as already stated in the seminal papers \cite{zurek,vedral}. Hence the following generalization is required
\begin{equation}
\varrho_{A|\Pi_j^B}\to\varrho_{A|E_j^B}=\mbox{Tr}_B(E_j^B\varrho_{AB}/p_j)
\end{equation}
where the elements of the POVM $E_j^B$ fulfill $\sum_j E_j^B={\mathbb 1}_B$
\begin{equation}
\label{eqdiscPOVM}
\delta_{A:B}(\varrho)=\min_{\{E_i^B\}}\left[S(\varrho_B)-S(\varrho)+S(A|\{E_i^B\})\right].
\end{equation}

Based on the convexity properties of the conditional entropy $S(A|\{E_i^B\})$ Hamieh et al. \cite{zaraket} (see
also \cite{datta}) show that POVM's which optimize discord are extremal or indecomposable, i.e.
they cannot be obtained by mixing other POVM's; further, it has been shown \cite{dariano} that extremal POVM's
for qubits are of rank 1 and can have between 2 and 4 elements.

\section{Unified picture through purification}
In \cite{winter} it was shown that \fer{given a mixed state} $\varrho_{AB}$ \fer{and its purification $|\psi_{ABC}\rangle$} through an ancilla qudit C, \fer{the following relation between the conditional entropy when B is measured,
and the entanglement of formation $E_F$ of the subsystem AC holds:}
\begin{equation}
\min_{\{E_i^B\}}S(A|\{E_i^B\})= E_F(\varrho_{AC})
\end{equation}
between the conditional entropy when B is measured, and the entanglement of formation $E_F$ of the subsystem AC. 
The minimization of the $AB$ conditional entropy over POVM measurements on B is thus equivalent to minimization of $E_F$ in $AC$ over all ensemble decompositions. Hence the number of elements giving the optimal ensemble decomposition of $E_F(\varrho_{AC})$ coincides with the number of elements of the POVM which minimizes $S(A|\{E_i^B\})$ \cite{winter}.\\

\textbf{Theorem 1}. Given a bipartite mixed state of two qubits $\rho_{AB}$ of rank 2, the optimal measurement giving the quantum discord is a 2 element POVM. The elements of such POVM are orthogonal projectors.

\textbf{Proof}. Consider a rank 2 state of two qubits with spectral decomposition $\rho_{AB}=\sum_{i=1}^{R=2}
\alpha_i|\psi_i\rangle\langle\psi_i|$. It\fer{s purification} by an ancilla qubit C \fer{has} the form
$|\Psi_{ABC}\rangle=\sum_{i=1}^{2} \sqrt{\alpha_i}|\psi_i\rangle|i\rangle_C$, where $|i\rangle_C$ forms an
orthonormal basis in the Hilbert space of the ancilla qubit. We can also Schmidt decompose this state as
$|\Psi_{ABC}\rangle=\sum_{i=1}^{m} \sqrt{\beta_i}|i\rangle_B|\phi_i\rangle_{AC}$, where $m=\min(d_B,d_Ad_C)=d_B=2$
with $d_B$ is the dimension of the Hilbert space of party B and so forth. So the partition AC has the form
$\rho_{AC}=\sum_{i=1}^{m=2} \beta_i|\phi_i\rangle\langle\phi_i|$ and hence is of rank 2. Wootters  \cite{wootters}
showed that the entanglement of formation of this two-qubit mixed state is obtained from an optimal decomposition
made up of {\it as many elements as its rank}, which in this case is 2, which in turn means that the POVM in B
that realizes such decomposition has 2 elements. Being optimal POVM's of rank 1, the 2 elements of such POVM are
necessarily orthogonal \cite{dariano}. This can be seen by noticing that a rank 1 POVM of 2 elements
$E_1=\alpha_1\ket{\phi_1}\bra{\phi_1}$ and $E_2=\alpha_2\ket{\phi_2}\bra{\phi_2}$ has to fulfill positivity
$E_i>0$ and normalization $E_1+E_2=\mathbb{1}$, which necessarily lead to orthogonality of its elements (this is
easy to show when the elements $E_i$ are written in Bloch form \cite{dariano}. QED \\

\textbf{Corollary}. The quantum discord of a rank 2 state of two qubits is given by
\begin{equation}
\delta_{A:B}(\varrho_{AB})=S(\varrho_B)-S(\varrho_{AB})+\mathcal{E}(C(\varrho_{AC})),
\end{equation}
with
\begin{equation}
\varrho_{AC}=\mbox{tr}_B\left(\sum_{i,j=1}^2\sqrt{\lambda_i\lambda_j}|\psi_i\rangle\langle\psi_j|\otimes|i\rangle_C  \langle j|\right),
\end{equation}
where $\{\lambda_i,|\psi_i\rangle\}$ is the spectral decomposition of $\varrho_{AB}$, and $|i_C\rangle$ is any orthonormal basis in $\mathcal{H}_C$. The function $\mathcal{E}$ is given by
\begin{equation}
\mathcal{E}(C)=h(\frac{1+\sqrt{1-C^2}}{2}),
\end{equation}
where
\begin{equation}
h(x)=-x\log_2x-(1-x)\log_2(1-x),
\end{equation}
and where $C(\rho)$ is the concurrence of $\rho$ \cite{wootters}, given by max$(0,l_1-l_2-l_3-l_4)$, with $l_i$ the eigenvalues of the hermitian matrix $R(\varrho_{AC})$, where $R(\rho)=\sqrt{\sqrt{\rho}\tilde{\rho}\sqrt{\rho}}$ and $\tilde{\rho}=(\sigma_y\otimes\sigma_y)\rho^*(\sigma_y\otimes\sigma_y)$.\\

Hence, to obtain the discord, diagonalize $\varrho_{AB}$ and use its eigenvalues and eigenvectors to construct $\varrho_{BC}$. Obtain the concurrence of $\varrho_{BC}$, substitute it in $\mathcal{E}(C)$ and obtain the optimal conditional entropy between A and B. This result was implicit in ref. \cite{chinos}.\\

\section{Quantum discord of rank 3 and 4 states}
%

Orthogonal measurements do not give the optimal discord for higher rank states, as first found in \cite{chinos2}
through a counterexample based on maximally discordant mixed
states (MDMS) \cite{MDMS}. Their study was limited to 3 element POVM's leading to a deviation of $\sim 2\times10^{-5}$ with respect
to orthogonal projectors. However we will give evidence that the \fer{set of} states where 3 and 4-element POVM's are needed
is indeed \fer{small} and the improvement in discord using Eq. (\ref{eqdiscPOVM}) is \fer{tiny}.

In the case of states $\varrho_{AB}$ with rank higher than 2, a purification  would
yield a qubit-qudit system in AC, whose optimal decomposition (for $E_F$) is not known. Therefore no analytical
tool can help us discriminate how many elements build the optimal POVM for the quantum discord of states with rank
higher than 2. However, it is known that for qubits the optimal measurements are given by rank 1 measurements with
2, 3 and 4 outcomes. This knowledge is based on two facts: i) the conditional entropy $S(A|\{E_i^B\})$ is a
concave function over the convex set of POVM's \cite{datta,zaraket}, hence only extremal POVM's will minimize it,
ii) extremal POVM's of qubits are rank 1 and have 2, 3 or 4 elements \cite{dariano}, i.e.
$E_j^B=\alpha_j\ket{\phi_j}\bra{\phi_j}$ ($j=2,...N$, $N\leq4$), with $\alpha$ real and nonzero and $\ket{\phi_j}$
are pure states (nonorthogonal unless $N=2$).

Even considering at most 4 elements POVM's,
the numerical analysis of this problem is challenging, as detailed in the following. We start parametrizing
the N elements of a POVM as:
\begin{eqnarray}
\tilde{E}_1&=&\alpha_1\ket{0}\bra{0}\nonumber\\
 \tilde{E}_j&=&\alpha_jU(\theta_j,\phi_j)\ket{0}\bra{0}U^\dagger(\theta_j,\phi_j)\ \ \ (1<j\le N),\nonumber
\end{eqnarray}
where $U(\theta,\phi)$ is a qubit rotation, plus a final global rotation $U(\Omega,\Phi)$
acting on all elements; i.e. $E_i=U(\Omega,\Phi)\tilde{E}_iU^\dagger(\Omega,\Phi)$. 
The completeness relation $\sum_i E_i={\mathbb 1}$ solves the coefficients in terms of the angles:
$\alpha_i=\alpha_i(\{\theta_j,\phi_j\})$.
This means running 6(8) loops in angles\footnote{\fer{Though for the case of $3$-element POVM's the number of loops can be reduced to 5 by using geometrical arguments.}}, for 3(4)-element POVM's, for each state,
in order to solve the minimization problem in the discord definition 
(\ref{eqdiscPOVM}). We must note that
orthogonal measurements are \fer{a limit case} in the definition of $3$-element POVM's; in the same way $4$ element POVM's
do not include the case of $3$ elements.

\begin{figure}[h!]
\includegraphics[width=8.8cm,height=10cm]{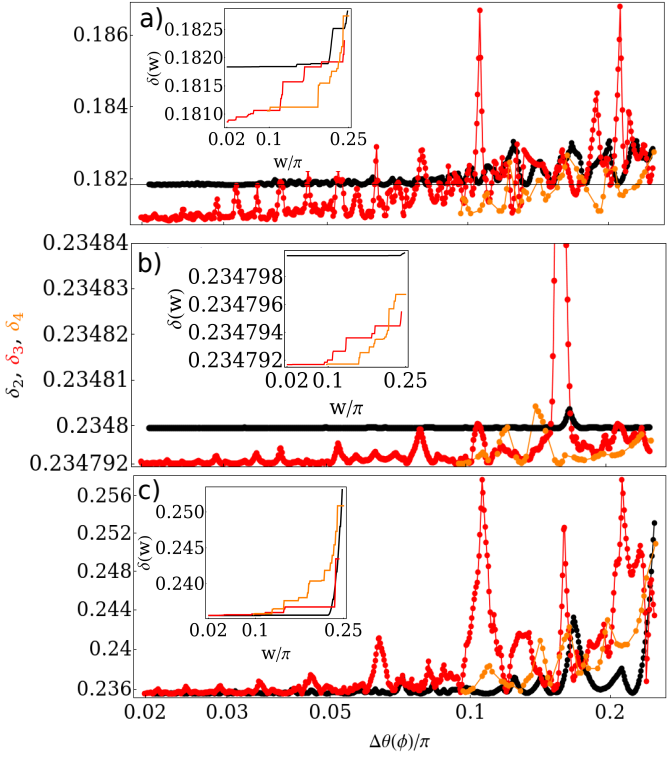}
\caption{(color online) Quantum discord minimized by 2 ($\delta_2$, black), 3($\delta_3$, red) and 4($\delta_4$, orange)
 elements POVM's with the given step size $\Delta\theta=\Delta\phi$ in the angles of the POVM elements $E_i$. The state\fer{s are} \fer{ a)
 the most deviant state we have found (highest point in fig.~\ref{fig3}), b) a MDMS of rank 3 with $m=0.11$, see \cite{MDMS}, and c) a random rank 3 state. In the inset we show the precision of $\delta_2$, $\delta_3$ and $\delta_4$, by plotting its minimum inside a box $\Delta\theta=\Delta\phi\in[\rm{w},0.25\pi]$. }}
\label{fig1}
\end{figure}

The numerical evaluation of discord is very sensitive to identification of the minimizing POVM and a proper 
scan of all possible POVM's requires small step sizes for the angles $\{\theta_j,\phi_j\}$.
 To give an idea on the sensitivity of the minimization on the step size in the angles, we show in
fig.~\ref{fig1} the values of discord for 2, 3 and 4-element POVM's
($\delta_2,\delta_3,\delta_4$) against the step size for \fer{three} states.
In \fer{all} cases it is surprising the oscillatory nature of
discord even for rather refined samplings (small angles). This demonstrates the importance to scan all POVM's 
over different step sizes to gather the
minimum, best approximating Eq.~(\ref{eqdiscPOVM}). We notice that a refinement until angular step sizes
$\sim 0.02\pi$ is feasible only for 2 and 3 elements POVM's. In any case, a good level of accuracy is obtained if
the lowest value of discord obtained \fer{among} different angular precisions is retained. \fer{The insets of fig.~\ref{fig1} show
the minimum value of discord obtained inside a box of angular precisions $\Delta\theta=\Delta\phi\in[\rm{w},0.25\pi]$.}

\begin{figure}
\includegraphics[width=7cm]{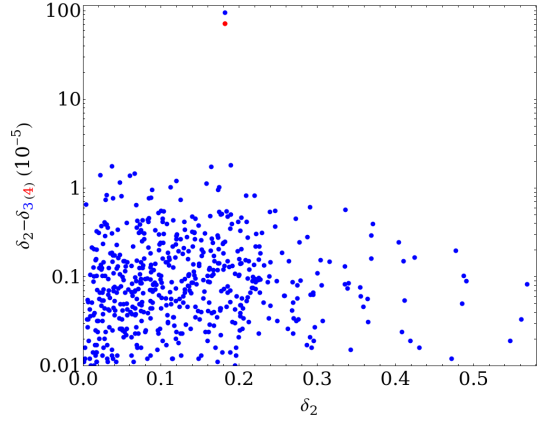}
\caption{\fer{(color online) Deviation of quantum discord $\delta_2-\delta_{3(4)}$ as given by 3- (blue) and 4-element (red) POVM's,
 for a scan of $10^5$ states of ranks 3 and 4. The density matrices are randomly generated following the Haar measure. The only state 
found with $\delta_2-\delta_4>0$ corresponds to that of fig.~\ref{fig1}a, with highest deviation.}}
\label{fig3}
\end{figure}

\fer{In fig.~\ref{fig1}a we show the state with highest deviation we have found (highest point in fig.~\ref{fig3}) in a scan of $10^5$ random states of rank 3 and 4, with $\delta_2-\delta_{3(4)}\sim10^{-3}$. This deviation is high above the typical deviation we have found of around $10^{-6}$.} In fig.~\ref{fig1}\fer{b} we show a 
MDMS of rank 3, separable, but with maximum discord versus classical
correlations \cite{MDMS}: $\varrho_{\mbox{\tiny MDMS}}=(1-\epsilon)(m\ket{00}\bra{00}+(1-m)\ket{11}\bra{11})+\epsilon\ket{\Psi^-}\bra{\Psi^-}$ with
$\ket{\Psi^-}=(\ket{01}-\ket{10})/\sqrt{2}$ the usual Bell state and values $\epsilon\fer{=0.2349602}$, $m=0.11$. In this case, POVM's with more than 2 elements are needed. This is actually a rather singular event, as we will see in fig.~\ref{fig3}. Indeed in
fig.~\ref{fig1}\fer{b} we show that an improvement of \fer{$\sim8\times10^{-6}$} is provided by 3,4-element POVM's
($\delta_{3(4)}$), as compared to orthogonal measurements ($\delta_2$). 
The most common situation is represented in figure \fer{~\ref{fig1}c}, for a generic 
state obtained by a random density matrix of rank 3.

Next we aim to establish the abundance of states for which 3 or 4 elements POVM's provide the improvement in
discord found in fig.~\ref{fig1}.
We then \fer{start} with a scan of Hilbert space where \fer{$10^5$} random density matrices of rank 3 and 4 have been
generated according to the Haar measure.  We plot in fig.~\ref{fig3} the deviations \fer{$\delta_{2}-\delta_{3(4)}$ (only when positive)}
versus the result for orthogonal measurements ($\delta_2$). \fer{In this figure }we have
sampled the angles from steps $\sim 0.3\pi$ until a lower limit of {\fer{w$=0.03\pi$ ($0.18\pi$)}} for 3 (4)-elements POVM's respectively. We
observe that the optimal discord is given by orthogonal POVM's, \fer{except for a $0.63\%$($0.001\%$) of states which have a typical deviation
of order $10^{-6}$ (see Table~\ref{tab.1}). In order to discriminate the dependence of such abundance and degree of deviations on the scan size ($N$), or the precision in the angles ($\Delta\theta=\Delta\phi\in[$w$,0.25\pi]$), we
 present in Table~\ref{tab.1} the results obtained for different sample characteristics $N,$w.

We find that for a given angular precision (for instance up to w$=0.03\pi$ in Table~\ref{tab.1}) the abundance of deviant states ($p$) is not sensitive to the sample size (the cases $N=3\times10^4$ and $10^5$ can be compared in the table). 
On the other hand discriminating states with $\delta_2>\delta_3$ in which these values are close, requires a quite refined angular precision (w), therefore leading to a  highly dependent abundance (p) (see strong variation of $p$ to find states with $\delta_2>\delta_3$ for $N=10^5$ in the table).
Due to computational limitations such a deep numerical study was only possible for 3-el. POVM's while for 4-el. POVM's we were only able to reach precisions up to w$=0.18\pi$, for which we have found only one deviating state. At the best precision reached we see that the typical (average) deviation $\delta_2-\delta_3$ is of the order $10^{-6}$ with a standard deviation of order $10^{-5}$.
}

\begin{table}\fer{
\caption{Probability ($p$) to find states with $\Delta\equiv\delta_2-\delta_{3(4)}>0$ in a sample of $N$ states whose discord is calculated with
angular precision up to w. We show the average deviation $\langle\Delta\rangle$ and its standard deviation $\sigma(\Delta)$. The standard
deviation is reduced by an order of magnitude if we remove the state a) of fig~\ref{fig1}. We do not show the average deviation for $\delta_2>\delta_4$ nor its standard 
deviation since only one state was found in the full scan.}
\label{tab.1}
\begin{center}

\begin{tabular}{|c|c|c|c|c|}
\cline{1-1}
$\delta_2>\delta_3$\\
\hline
$N$& w$/\pi$ & $p$ & $\langle\Delta\rangle$ & $\sigma(\Delta)$\\
\hline
$3\cdot 10^4$& $0.03$ & $6.6\cdot10^{-3}$ & $1.6\cdot10^{-6}$ & $2.6\cdot10^{-6}$\\
\hline
$10^5$& $0.05$ & $5\cdot10^{-4}$ & $2\cdot10^{-5}$ & $1.2\cdot10^{-4}$\\
\hline
$10^5$& $0.03$ & $6.3\cdot10^{-3}$ & $3\cdot10^{-6}$ & $3.7\cdot10^{-5}$\\
\hline
$10^5$& $0.025$ & $8\cdot10^{-3}$ & $2.8\cdot10^{-6}$ & $3.4\cdot10^{-5}$\\
\hline
$10^5$& $0.02$ & $1.4\cdot10^{-2}$ & $2.1\cdot10^{-6}$ & $2.6\cdot10^{-5}$\\
\hline
%
%

$\delta_2>\delta_4$\\
\hline
$3\cdot10^4$ & $0.2$ & $0$ & -- & --\\
\hline
$10^5$ & $0.2$ & $10^{-5}$& -- & --\\
\hline
$10^5$ & $0.18$ & $10^{-5}$& -- & --\\
\hline
\end{tabular}
\end{center}}
\end{table}

The set of states for which we find improvements with $\delta_{3(4)}$ 
is rather small. An interesting question is whether these states lie in the neighborhood of rank 3 MDMS 
\cite{MDMS,chinos2} or are distributed everywhere in the $\{\mathcal{J},\mathcal{I}\}$ diagram (figure 1 in 
the same reference). \fer{We first stress
that the state in fig.~\ref{fig1}a is nowhere near the MDMS border (it has $\{\mathcal{J}\simeq0.17,\mathcal{I}\simeq0.1809\}$).}
\fer{Moreover} we can gain some insight about the smallness of \fer{the MDMS} neighborhood by investigating the state of rank 3 in fig.~\ref{fig1}\fer{b} \cite{MDMS} whereby we
perturbate it with a Bell state. That is we study the state $\varrho=(1-\lambda)\varrho_{\mbox{\tiny MDMS}}+\lambda\ket{\Phi^+}\bra{\Phi^+}$ with
$\ket{\Phi^+}=(\ket{00}+\ket{11})/\sqrt{2}$. The effect of this
perturbation is to move the state away from the   border in the $\{\mathcal{J},\mathcal{I}\}$ diagram. We find
that already for \fer{$\lambda\simeq0.002$} we reach the transition where $\delta_2=\delta_3=\delta_4$, meaning that, at
least in the neighborhood of the MDMS border, 3,4-element POVM's are needed in a very tiny region. 

Summarizing our numerical analysis for randomly generated states, we have found that orthogonal measurements
are $almost$ enough \fer{in the following sense}. The improvements
$\delta_{3(4)}<\delta_2$ do occur but \fer{they represent small corrections (a maximum deviation $10^{-3}$ was found for only one state in a sample of $10^5$ while other deviations were up to $10^{-5}$, fig.~\ref{fig3}) and} they can
be numerically \fer{appreciated} only for a tiny set of states \fer{(with the mentioned minimization up to precision w$=0.02\pi$ these states appear with probability $\sim10^{-2}$ for $\delta_3$, while for $\delta_4$ they appear with probability $\sim10^{-5}$ at w$=0.18\pi$).}

Given this numerical evidence we give the following upper bound:

\section{Observation}
The entanglement of formation of a $2\otimes N$ rank 2 bipartite state $\varrho_{BC}$ has a {\it tight upper bound} given by an optimal decomposition 
of two elements:
\begin{equation}
E_F(\varrho_{BC})\leq\min_{\{p_k,|\phi_{BC}^k\rangle\}}\sum_{k=1}^{\bf 2}p_kE(|\phi_{BC}^k\rangle),
\end{equation}
with $E(|\phi_{BC}^k\rangle)=S(\mbox{tr}_B(\ket{\phi_{BC}^k}\bra{\phi_{BC}^k}))$. \fer{The deviation from the equality 
is on average of order $10^{-6}$}.\\

To show this, take the $2\otimes N$, rank 2, state in spectral form $\varrho_{BC}=\sum_{k=1}^2\lambda_k|\psi_{BC}^k\rangle\langle\psi_{BC}^k|$. It can be purified by an ancilla qubit as $|\psi_{ABC}\rangle=\sum_{k=1}^2\sqrt{\lambda_k}|e_A^k\rangle|\psi_{BC}^k\rangle$, with $|e_A^k\rangle$ any orthonormal basis in $\mathcal{H}_A$. Again according to \cite{winter}:
\begin{equation}
\label{eq8}
E_F(\varrho_{BC})=\min_{\{E_j^A\}}S(\{E_j^A\}|B)\leq\min_{\{\Pi_j^A\}}S(\{\Pi_j^A\}|B),
\end{equation}
where restriction to orthogonal measurements 'spoils' the minimization. Given the numerical evidence we have provided, the improvement 
of doing full minimization is \fer{on average} at the level of \fer{$10^{-6}$ (see Table~\ref{tab.1})}.

The conditional entropy for a given orthonormal measurement $\Pi_k^A=|\xi^k\rangle\langle\xi^k|$ is given by
\begin{equation}
\label{eq9}
S(\{\Pi_k^A\}|B)=\sum_{k=1}^2 p_k S(\rho_B^k)=\sum_{k=1}^2 p_k E(\rho_{BC}^k)
\end{equation}
with $p_k=\mbox{tr}(\Pi_k^A\varrho_{ABC})$ and $\rho_B^k=\mbox{tr}_{AC}(\Pi_k^A\varrho_{ABC})/p_k$. In the last equality we have used the fact that $\rho_{BC}^k$ is pure. This can be seen by writing explicitly
\begin{eqnarray}
\rho_{BC}^k&=&\mbox{tr}_A(\Pi_k^A\varrho_{ABC})=\nonumber\\
&=&\sum_{i,j=1}^2e_i(k)e_j^*(k)\sqrt{\lambda_i\lambda_j}|\psi_{BC}^i\rangle\langle\psi_{BC}^j|=\nonumber\\
&=&|\phi_{BC}^i\rangle\langle\phi_{BC}^i|
\end{eqnarray}
with $|\phi_{BC}^i\rangle=\sum_{i=1}^2e_i(k)\sqrt{\lambda_i}|\psi_{BC}^i\rangle$, and $e_i(k)=\langle\xi^k|e_i\rangle$. So, finally, $S(\rho_B^k)=E(|\phi_{BC}^k\rangle)$. QED \\

We stress the fact that instead of minimization over ensemble decompositions with a number of elements ranging from $R$ to $R^2$ ($R$ is the rank of the state), as shown to be sufficient by Uhlmann \cite{uhlmann}, we can safely restrict to decompositions with 2 elements if we are not interested in states which are rare to find and have deviations which are probably quite small, as we have seen when perturbing a MDMS extremal state.

\section{Bloch formula for quantum discord}

We finish by giving an alternative formula for the computation of quantum discord of generic two-qubit states of any rank. Writing the POVM elements in Bloch form
\begin{equation}
\label{POVMbloch}
E_i^A=\alpha_i({\mathbb{1}}_A+\vec{n}_i\cdot\vec{\sigma}_A),
\end{equation}
with the positivity and normalization (completeness) conditions
\begin{equation}
\alpha_i>0\ ,\ \sum_i\alpha_i=1\ ,\ \sum_i\alpha_i\vec{n}_i=\vec{0}\ ,\ (|\vec{n}_i|=1),
\end{equation}
and the density matrix $\varrho_{AB}$ (coming from purification of $\varrho_{BC}$ plus tracing subsystem C) given also in Bloch form
\begin{equation}
\label{rhoBloch}
\varrho_{AB}=\frac{1}{4}\left(\mathbb{1}_{AB}+\vec{a}\cdot\vec{\sigma}_A+\vec{b}\cdot\vec{\sigma}_B+\sum_ic_i\sigma_A^i\otimes\sigma_B^i        \right)
\end{equation}

\begin{widetext}
\begin{eqnarray}
\label{alternDisc1}
\delta_{A:B}(\varrho_{AB})&=&S_B-S_{AB}+\min_{\{\alpha_i,\vec{n}_i\}}\sum_{i=1}^m\alpha_i(1+\vec{n}_i\cdot\vec{b})\sum_\pm H(\lambda_i^\pm(\vec{a},\vec{b},\vec{cn}_i))\\
\label{alternDisc2}
\lambda_i^\pm(\vec{a},\vec{b},\vec{cn}_i)&=&\frac{1}{2}\left(1\pm\left|\frac{\vec{a}+\vec{cn}_i}{1+\vec{b}\cdot\vec{n}_i}\right|\right)\ ,\ \vec{cn}_i=\{c_xn_{i,x},c_yn_{i,y},c_zn_{i,z}\}\\
\label{alternDisc3}
S_B&=&\sum_\pm H(\frac{1}{2}(1\pm|\vec{b}|))\\
\label{alternDisc4}
S_{AB}&=&S(\varrho_{AB})
\end{eqnarray}
\end{widetext}

(notice that when we write $\vec{a}\cdot\vec{\sigma}_A$, we mean $\vec{a}\cdot\vec{\sigma}_A\otimes\mathbb{1}_B$), we enunciate the following theorem:\\

\textbf{Proposition}. The discord $\delta_{A:B}(\varrho_{AB})$ with $\varrho_{AB}$ of {\it any rank}, written as in eq. (\ref{rhoBloch}), is given by eqs.(\ref{alternDisc1}-\ref{alternDisc4}) and $m(=2,3,4)$ is the number of elements of the extremal POVM.\\

The minimization is restricted by the conditions for {\it extremality} \cite{dariano} for each number of POVM elements (m):
\begin{eqnarray}
\label{extr2}
&\mbox{m=2:}&\mbox{  all extremal, from normalization they follow:}\nonumber\\
&& \ \alpha_i=\frac{1}{2}\ ,\ \vec{n}_1=-\vec{n}_2\equiv\vec{n}\mbox{  (i.e. orthogonal)}\\
\label{extr3}
&\mbox{m=3:}&\mbox{  all extremal, from normalization they follow:}\nonumber\\
&& \ \alpha_3=1-\alpha_1-\alpha_2\ ,\ \vec{n}_3=-\frac{1}{\alpha_3}(\vec{n}_1+\vec{n}_2)\\
\label{extr4}
&\mbox{m=4:}&\ \ \vec{n}_i\mbox{ not in the same plane;}\nonumber\\
&& \ \mbox{normalization yields  }\sum_{i=1}^4\alpha_i\vec{n}_i=\vec{0}
\end{eqnarray}

\textbf{Proof}. Simple algebra yields the probabilities and outcomes of each measurement:
\begin{eqnarray}
p_k&\equiv&\mbox{tr}\left(E_i^B\varrho_{AB}\right)=\alpha_i(1+\vec{n}_i\cdot\vec{b})\\
\rho_A^k&\equiv&\mbox{tr}_{B}\left(E_i^B\varrho_{AB}\right)/p_k=\nonumber\\
&=&\frac{1}{2}\left(\mathbb{1}_A+\frac{(\vec{a}+\vec{cn}_k)\cdot\vec{\sigma}_A}{1+\vec{b}\cdot\vec{n}_k}\right)
\end{eqnarray}
whose entropy can be calculated from the eigenvalues of $\rho_A^k$ ($\lambda_k^\pm(\vec{a},\vec{b},\vec{cn}_k)$) as defined in eq. (\ref{alternDisc2}). The condition that for $m=4$ extremality implies that the POVM elements cannot lie in the same plane, was derived in \cite{dariano}, as well as the normalization conditions.\\ 

In summary, we have proven that the quantum discord of rank 2 two-qubit mixed states is obtained {\it using only orthogonal projectors} as measurements.
 Strong numerical evidence has been given to conjecture that they are almost sufficient for higher ranks, except for states which appear with
 probability \fer{$\sim10^{-2}$} and have negligible deviations \fer{on average }of order \fer{$10^{-6}$ in a sample of $10^5$ states (w$=0.02\pi$ for $\delta_3$).
We discussed the importance of the states sample size and of the precision in scanning all possible measurements, showing the need of using a minimization procedure over different angular step sizes. Two} example\fer{s} w\fer{ere} given
 \fer{in figs.~\ref{fig1}a and \ref{fig1}b,} \fer{both of rank 3}, showing that 3 and
 4 element POVM's give a better quantum discord, though with a very small improvement \fer{(a maximum of order $10^{-3}$)}. Based on the connection \cite{winter} between conditional entropy and entanglement of formation we have given
 a related tight upper bound, namely that the entanglement of formation of a $2\otimes N$ system of rank 2 is obtained by 2 element decompositions
 with very high probability and precision. Finally, an alternative quantum discord formula for generic mixed states of two qubits was given in terms
 of the Bloch vectors of the state, where minimization is performed over the Bloch form of POVM's elements.
 \fer{As a side remark, we note that Gaussian discord in continuous variable systems is also an example where projective, but not orthogonal
 measurements are optimal \cite{Paris,Adesso}.}

\acknowledgments 
Funding from FISICOS (FIS2007-60327), Govern Balear (AAEE0113/09) and CoQuSys (200450E566) projects, JaeDoc (CSIC) and
Juan de la Cierva program are acknowledged. \fer{We also thank Matthias Lang for useful comments. Numerical calculations have
been performed using the IFISC and GRID-CSIC cluster facilities (Ref. 200450E494).}


%

%



\begin{thebibliography}{0}
\bibitem{info-no-ent}\Name{Knill E. \and Laflamme R.}\REVIEW{Phys. Rev. Lett.}{81}{1998}{5672}; \Name{Braunstein S. L., Caves C. M, Jozsa R., Linden N., Popescu S. \and Schack R.}\REVIEW{Phys. Rev. Lett.}{83}{1999}{1054};
\Name{Meyer D. A.}\REVIEW{Phys. Rev. Lett.}{85}{2000}{2014};
\Name{E. Biham, Brassard G., Kenigsberg D. \and Mor T. T.}\REVIEW{Theor. Comput. Sci.}{320}{2004}{15};
\Name{A.Datta, Shaji A. \and Caves C. M.}\REVIEW{Phys. Rev. Lett.}{100}{2008}{050502};
\Name{Lanyon B. P., Barbieri M., Almeida M. P. \and White A. G.}\REVIEW{Phys. Rev. Lett.}{101}{2008}{200501};
\Name{Vedral V.}\REVIEW{Found. Phys.}{40}{2010}{1141}.
\bibitem{zurek}\Name{Ollivier H. \and Zurek W.  H.}\REVIEW{Phys. Rev. Lett.}{88}{2001}{017901}.
\bibitem{vedral}\Name{Henderson L. \and Vedral V.}\REVIEW{J. Phys. A}{34}{2001}{6899}.
\bibitem{maziero}\Name{Maziero J., C\'eleri L. C., Serra R. M. \and Vedral V.}\REVIEW{Phys. Rev. A}{80}{2009}{044102}.
\bibitem{luo2008}\Name{Luo S.}\REVIEW{Phys. Rev. A}{77}{2008}{042303}.
\bibitem{mazhar2010}\Name{Ali M., Rau A. R. P. \and Alber G.}\REVIEW{Phys. Rev. A}{81}{2010}{042105}.
\bibitem{mazzola2010}\Name{Mazzola L., Piilo J. \and Maniscalco S.}\REVIEW{Phys. Rev. Lett.}{104}{2010}{200401}.
\bibitem{guo}\Name{Xu J. S., Xu X. Y., Li C. F., Zhang C. J., Zou X. B. \and Guo G. C.}\REVIEW{Nat. Commun.}{1}{2010}{7}.
\bibitem{zaraket}\Name{Hamieh S., Kobes R. \and Zaraket H.}\REVIEW{Phys. Rev. A}{70}{2004}{052325}.
\bibitem{dariano}\Name{D'Ariano G. M., Presti P. L. \and Perinotti P.}\REVIEW{J. Phys. A: Math. Gen.}{38}{2005}{5979}.
\bibitem{winter}\Name{Koashi M. \and Winter A.}\REVIEW{Phys. Rev. A}{69}{2004}{022309}.
\bibitem{datta}\Name{Datta A.}\REVIEW{arXiv:0807.4490}{}{2008}{}.
\bibitem{wootters}\Name{Wootters W. K.}\REVIEW{Phys. Rev. Lett.}{80}{1998}{2245}.
\bibitem{chinos}\Name{Cen L., Li X., Shao J. \and Yan Y.}\REVIEW{Phys. Rev. A}{83}{2011}{054101}.
\bibitem{chinos2}\Name{Chen Q., Zhang C., Yu S., Yi X.X. \and Oh C.H.}\REVIEW{arXiv:1102.0181}{}{2011}{}.
\bibitem{MDMS}\Name{Galve F., Giorgi G. L. \and Zambrini R.}\REVIEW{Phys. Rev. A}{83}{2011}{012102}.
\bibitem{uhlmann}\Name{Uhlmann A.}\REVIEW{Open Syst. Inf. Dyn.}{5}{1998}{209}.
\bibitem{Paris}\Name{Adesso G. \and Datta A.}\REVIEW{Phys. Rev. Lett.}{105}{2010}{030501}.
\bibitem{Adesso}\Name{Giorda P. \and Paris M.}\REVIEW{Phys. Rev. Lett.}{105}{2010}{020503}.


\end{thebibliography}
\end{document}